\documentclass[aps,onecolumn,preprintnumbers,showpacs,showkeys,nofootinbib,
superscriptaddress  
]{revtex4}

\usepackage{multirow}
\usepackage{epsfig}
\usepackage{amssymb,amsmath,amsfonts,amsthm,graphicx,psfrag}
\usepackage{hyperref}
\newcommand{\beq}{\begin{eqnarray}}
\newcommand{\eeq}{\end{eqnarray}}

\newcommand{\tr}{\operatorname{Tr}}

\begin{document}
\preprint{}

\title{
Temperature dependence of shear viscosity of $SU(3)$--gluodynamics within lattice simulation
}

\author{N.~Yu.~Astrakhantsev}
\email[]{nikita.astrakhantsev@itep.ru}
\affiliation{Institute for Theoretical and Experimental Physics, Moscow, 117218 Russia\\
and Moscow Institute of Physics and Technology, Dolgoprudny, 141700 Russia}

\author{V.~V.~Braguta}
\email[]{braguta@itep.ru}
\affiliation{Institute for High Energy Physics NRC "Kurchatov Institute", Protvino, 142281 Russian Federation\\
Far Eastern Federal University, School of Biomedicine, 690950 Vladivostok, Russia \\
and Institute for Theoretical and Experimental Physics, Moscow, 117218 Russia \\ 
and Moscow Institute of Physics and Technology, Dolgoprudny, 141700 Russia
}

\author{A.~Yu.~Kotov}
\email[]{kotov@itep.ru}
\affiliation{Institute for Theoretical and Experimental Physics, Moscow, 117218 Russia}

\begin{abstract}
In this paper we study the shear viscosity temperature dependence of $SU(3)$--gluodynamics within lattice simulation. To do so, we measure the correlation functions of energy-momentum tensor in the range of temperatures $T/T_c\in [0.9, 1.5]$.
To extract the values of shear viscosity we used two approaches. The first one
is to fit the lattice data with some physically motivated ansatz for the spectral function 
with unknown parameters and then determine shear viscosity. The second approach
is to apply the Backus-Gilbert method which allows to extract shear viscosity from the lattice data nonparametrically.  
The results obtained within both approaches agree with each other. Our results 
allow us to conclude that within the temperature range $T/T_c \in [0.9, 1.5]$ SU(3)--gluodynamics reveals the properties of a strongly interacting system, which cannot be described perturbatively, and has the ratio $\eta/s$ close to the value ${1}/{4\pi}$ in $N = 4$ Supersymmetric Yang-Mills theory.
\end{abstract}



\maketitle

\subsection*{Introduction}
Modern heavy ion collisions experiments such as RHIC and LHC are aimed at study of quark-gluon plasma (QGP). Hydrodynamic description of QGP evolution proved to be efficient in understanding of experimental results \cite{Kolb:2003dz, Ollitrault:2008zz}. 
Despite this success, hydrodynamics is only an effective theory which correctly represents
dynamics of infrared degrees of freedom. Parameters of this effective theory, such as shear viscosity, 
bulk viscosity, conductivity, etc. cannot be calculated within hydrodynamics itself but
must be determined either from the experiment or from calculation based on the 
first principles.  

The measurement of elliptic flow \cite{Ackermann:2000tr, Adler:2002pu} allows to estimate the value of  
QGP shear viscosity. In particular, hydrodynamic approximation describes the
experimental data if the ratio of shear viscosity $\eta$ to entropy density $s$ lies within the range 
$\eta/s = (1-2.5)\times 1/4\pi$ \cite{Song:2012ua}. This value is close to the result of $N = 4$ 
Supersymmetric Yang-Mills (SYM) theory at strong coupling $\eta/s = 1/4\pi$ \cite{Policastro:2001yc} and deviates from the calculation at weak coupling $\eta/s \sim \mbox{const}/(g^4 \log(1/g)) \sim 1$ 
\cite{Arnold:2000dr,Arnold:2003zc}.    
 
From this consideration one can conclude that the small value of ratio $\eta/s$ is governed by nonperturbative dynamics. Attempts of nonperturbative calculation of shear viscosity were undertaken
in papers \cite{Ozvenchuk:2012kh, Marty:2013ita, Berrehrah:2016vzw, Christiansen:2014ypa}.
Unfortunately it is rather difficult to estimate the systematic uncertainty of these 
approaches, so today the analytical approch which
fully accounts for nonperturbative dynamics of QGP based on the first principles is absent. 
For this reason the only way to calculate shear viscosity of QGP is the lattice simulation of QCD.  

Despite considerable progress in the lattice study of QCD properties today it is not 
possible to calculate shear viscosity of QGP with dynamical quarks. Even shear viscosity study within pure gluodynamics is an extremely complicated task. 
There are only few attempts to calculate shear viscosity of SU(3)--gluodynamics undertaken in papers \cite{Karsch:1986cq, Nakamura:2004sy, Meyer:2007ic, Meyer:2009jp, Mages:2015rea} 
and SU(2)--gluodynamics undertaken in papers \cite{Braguta:2013sqa, Astrakhantsev:2015jta}. 
In this paper we are going to study the temperature dependence of shear viscosity
of SU(3)--gluodynamics in the vicinity of the confinement/deconfinement phase transition $T/T_c \in [0.9, 1.5]$. 

The paper is organized as follows. In the next section we describe the details of the calculation. 
Section III is devoted to the calculation of shear viscosity from lattice measurements of 
the energy-momentum tensor correlation function. 
In the last section our results are discussed and the conclusion is drawn.  

\subsection*{Details of the calculation}

Shear viscosity is related to the Euclidean correlation function of the energy-momentum tensor 
$T_{\mu \nu} = \frac 1 4 \delta_{\mu \nu} F_{\alpha \beta}^a F_{\alpha \beta}^a - F_{\mu \alpha }^a F_{\nu \alpha}^a$ (here for simplicity we omitted the trace anomaly):
\begin{equation}\begin{split}
	C(x_0)=T^{-5} \int d^3{\textbf x}\langle T_{12}(0)T_{12}(x_0,{\textbf x})\rangle,
\label{correlator}
\end{split}\end{equation}
where $T$ is the temperature of the system.
The correlation function (\ref{correlator}) can be written in terms of the spectral function $\rho(\omega)$ as follows

\begin{equation}\begin{split}
	C(x_0)=T^{-5} \int_{0}^{\infty}\rho(\omega)\frac{\cosh \omega(\frac1{2T}-x_0)}{\sinh\frac{\omega }{2T}} d\omega.
\label{spectr_corr}
\end{split}
\end{equation}

The spectral function contains valuable information about the properties of a medium. To find shear viscosity from the spectral function one uses the Kubo formula \cite{Kubo:1957mj} 
\begin{equation}\begin{split}
	\eta=\pi\lim\limits_{\omega\to0} \frac{\rho(\omega)}{\omega},
\end{split}\end{equation}
Lattice calculation of shear viscosity can be divided into two parts. The first part is the measurement of the correlation function $C(x_0)$ with sufficient accuracy. 
This part of the calculation requires large computational resources but for the gluodynamics the accuracy of the correlator can be dramatically improved with the help of the two-level algorithm \cite{Meyer:2002cd}.
The second part is the determination 
of the spectral function $\rho(\omega)$ from the correlation function $C(x_0)$. The last part of 
the calculation is probably the most complicated, since one should determine continuous spectral
 function $\rho(\omega)$ from the integral equation (\ref{spectr_corr})
for the set of $\mathcal{O}(10)$ values of the function $C(x_0)$ measured within the lattice simulation. 

Below we will need the the properties of the spectral function. 
First we recall very general properties: the positivity of the spectral function $\rho(\omega)/\omega \geqslant 0$ and oddness: $\rho(-\omega) = - \rho( \omega)$. At large frequencies one expects that the asymptotic freedom manifests itself
in the approach of the real spectral function to the one calculated at weak coupling. 
For this reason it is also important to write the expression for the spectral function 
in the leading-order approximation in the strong coupling constant \cite{Meyer:2008gt}
\begin{equation}
\rho^{LO} (\omega) = \frac 1 {10} \frac {d_A} {(4 \pi)^2} \frac {\omega^4} {\tanh (\frac {\omega} {4T})} + \biggl ( \frac {2 \pi} {15} \biggr )^2 d_A T^4 ~ \omega \delta(\omega),
\label{rho_tree_level}
\end{equation}
where $d_A=N_c^2-1=8$ for the SU(3)--gluodynamics.

The next-to-leading order expression for the 
spectral function at a large $\omega$ is also known \cite{Kataev:1981gr}:
\begin{equation}
\lim_{\omega \to \infty} \rho^{NLO} (\omega) = \frac 1 {10} \frac {d_A} {(4 \pi)^2} \omega^4 \biggr (1 - \frac {5 \alpha_s N_c} {9 \pi}  \biggl )
\label{rho_NLO}
\end{equation}
It should be noted here that at a large $\omega$ the spectral function scales as $\rho(\omega) \sim \omega^4$,
what leads to a large perturbative contribution to the correlation function for all values of the Euclidean time $x_0$.
Calculation shows that even at the $x_0 = 1/(2T)$ the tree level contribution is $\sim 80-90\%$ of the total value of the 
correlation function. Note also that the large $\omega$ behavior of the spectral function 
leads to a fast decrease of the correlation function $C(x_0) \sim 1/x_0^5$ for small $x_0$. 
For this reason the signal/noise ratio for the $C(x_0)$ is small at $x_0 \gg a$
and the lattice measurement of the correlation function at $x_0 \sim 1/(2T)$ becomes computationally very expensive.

In numerical simulation we use the Wilson gauge action for the SU(3)--gluodynamics
\begin{equation}\begin{split}
	S_g=\beta\sum_{x,\mu<\nu}\left(1-\frac1{3} \mbox{Re} \tr U_{\mu,\nu}(x)\right),
\end{split}\end{equation}
where $U_{\mu,\nu}(x)$ is the product of the link variables along the elementary rectangular $(\mu, \nu)$, which starts at $x$.

For the tensor $F_{\mu\nu}$ we use the clover discretization scheme:
\begin{equation}\begin{split}
	F^{(clov)}_{\mu\nu}(x)=\frac1{4iga^2}(V_{\mu,\nu}(x)+V_{\nu,-\mu}(x) 
	+ V_{-\mu,-\nu}(x)+V_{-\nu,\mu}(x) ),\\
	V_{\mu,\nu}(x)=\frac12(U_{\mu,\nu}(x)-U_{\nu,\mu}(x)).
\end{split}
\label{clover}
\end{equation}
One can easily define the energy-momentum tensor on the lattice using its continuum expression and lattice discretization (\ref{clover}) for the $F_{\mu\nu}$ tensor. 

Note also
that instead of the correlation function $\langle T_{12}(x)T_{12}(y)\rangle$ in this paper we measure the correlation function 
$\frac12(\langle T_{11}(x)T_{11}(y)\rangle-\langle T_{11}(x)T_{22}(y)\rangle)$. Both correlation functions are equal in 
the continuum limit \cite{Karsch:1986cq}. Meanwhile the renormalization properties of the diagonal components of energy-momentum tensor $T_{\mu\nu}$ are known (see below).

It has become conventional to present the value of shear viscosity as the ratio viscosity-to-entropy-density $\eta/s$. For homogeneous systems the entropy density $s$ can be expressed as $s = \frac{\epsilon+p}T$, where $\epsilon$ is the energy density and $p$ is the pressure. These thermodynamic quantities were 
measured with the method described in \cite{Engels:1999tk}. 	

The energy-momentum tensor in the continuum theory is a set of Noether currents which are related to the translation invariance of the action.  In the lattice formulation of field theory continuum rotational invariance does not exist 
and the renormalization for energy-momentum tensor is required. 
For the correlation function considered in this paper the renormalization is multiplicative \cite{Meyer:2011gj}. 
The renormalization factors depend on the discretization scheme. For instance, for the diagonal component of 
$T_{\mu\nu}$ (when $\mu=\nu$) and the plaquette-based discretization of $T_{\mu\nu}$: 
$\displaystyle T_{\mu\mu}=\frac{2}{a^4g^2}\left(-\sum\limits_{\nu\ne\mu}\tr U_{\mu,\nu}(x)+\sum\limits_{\nu,\sigma\ne\mu,\sigma>\nu}\tr U_{\sigma,\nu}(x)\right)$ the renormalization factors are related to the anisotropy coefficients \cite{Engels:1999tk,Meyer:2007fc}: $T^{(ren)}_{\mu\nu}=Z^{(plaq)}T_{\mu\nu}^{(plaq)}$, $\displaystyle Z^{(plaq)}=1-\frac12g_0^2(c_{\sigma}-c_{\tau})$, where $c_{\sigma}$ and $c_{\tau}$ are defined in \cite{Engels:1999tk}.
 
 Using the renormalization factors for the plaquette-based discretization of $T_{00}$, we can find the renormalization factors for the clover discretization simply by fitting the vacuum expectation values of the renormalized $T_{00}$: 
$Z^{(plaq)}\langle T_{00}^{(plaq)}\rangle=Z^{(clov)}\langle T_{00}^{(clov)}\rangle$.
 
\begin{figure}[t]
\begin{center}
\includegraphics[scale = 0.5]{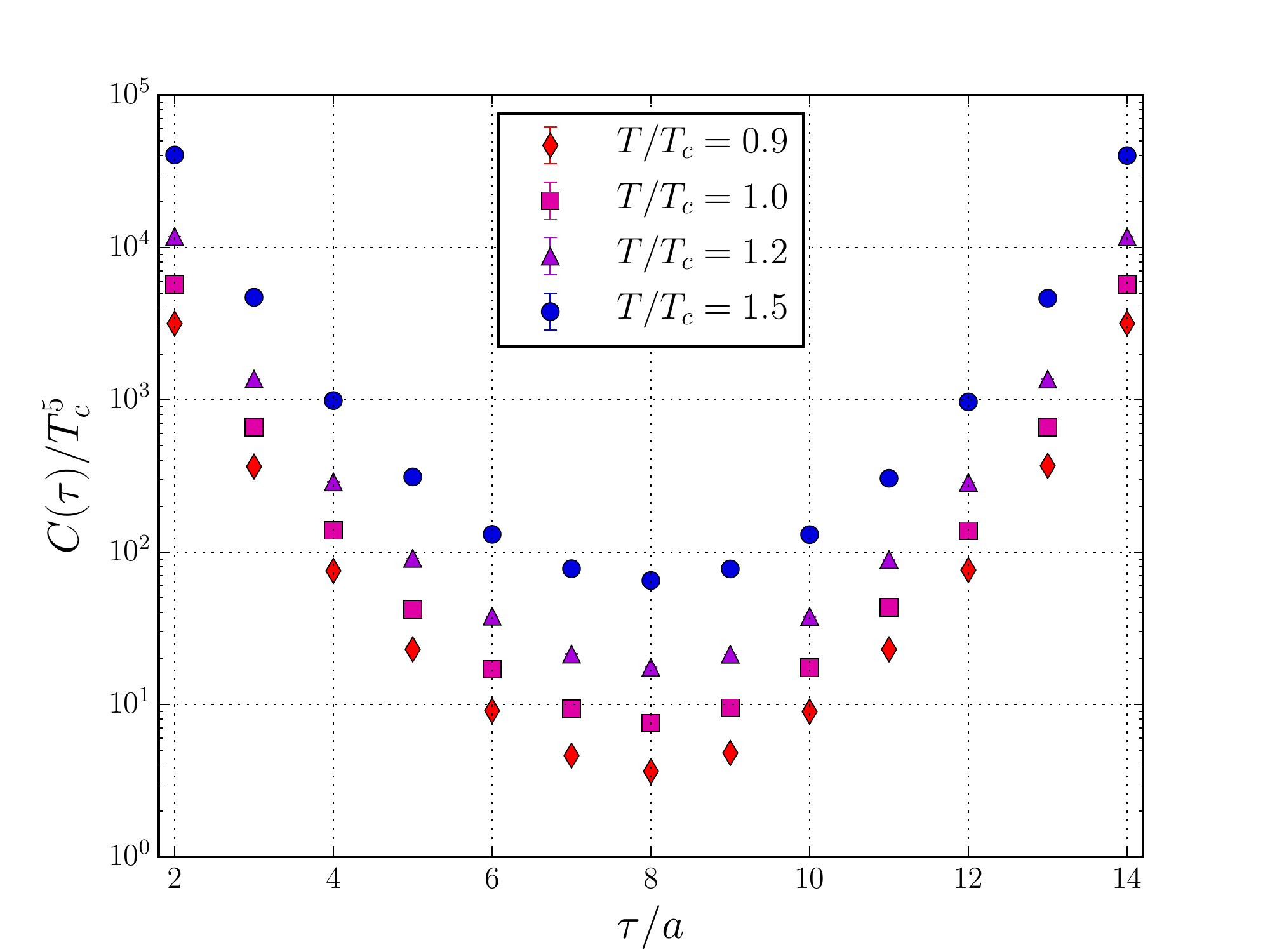}
\caption{The correlation functions $C(x_0)$ for the temperatures $T/T_c=0.9, 1.0, 1.2, 1.5$.}
\label{fig:correlation_f}
\end{center}
\end{figure}

\section*{Numerical results}
\subsection*{The spectral function from the fitting procedure.}

  We measured the correlation functions $C(x_0)$ on the lattice $16\times32^3$ with the following set of the $\beta$--parameter: 
$\beta=6.491, 6.512, 6.532, 6.575, 6.647, 6.712, 6.811, 6.897$, 
which correspond to the temperatures $T/T_c\simeq0.9, 0.925, 0.95, 1.0, 1.1, 1.2, 1.35, 1.5$. Application of two-level algorithm allowed us to get 
uncertainties not larger than $\sim 2-3\%$ at the distance $T x_0 = 0.5$ for all temperatures under 
consideration. For the other points the accuracy is even better. In Fig.~\ref{fig:correlation_f} we 
plot the correlation functions for various temperatures.

  The next step in the calculation of shear viscosity is the extraction of the spectral function from 
the integral equation (\ref{spectr_corr}).  In this section we are going to use
physically motivated ansatzes for the spectral function with unknown parameters. These parameters will be determined 
through the fitting procedure.
 
The first anzatz which will be used in the calculation is motivated by QCD sum rules \cite{Shifman:1978bx}. In order to build a tentative spectral function
we join the first-order hydrodynamic behavior at small frequencies with the asymptotic freedom at large frequencies
\footnote{Note that the frequency $\omega$ is measured in physical units.}
\begin{equation}
\rho_1(\omega) = B T^3 \omega  \theta(\omega_0 - \omega) + A  \rho_{lat} ( \omega )  \theta(\omega-\omega_0).
\label{rho1}
\end{equation}
In the last formula $\rho_{lat}(\omega)$ is the tree level lattice expression for the spectral function
calculated for the correlation function 
$\sim \frac12(\langle T_{11}(x)T_{11}(y)\rangle-\langle T_{11}(x)T_{22}(y)\rangle)$ with clover discretization
of the tensor $F_{\mu \nu}$ at lattice with the fixed $L_t$ and $L_s \to \infty$. The function $\rho_{lat}(\omega)$
was calculated in the paper \cite{Astrakhantsev:2015jta}. 

It was noted above that the correlation function $C(x_0)$ is very sensitive to the ultraviolet 
properties of the spectral function. To get accurate description of the lattice data
we used the lattice spectral function at large frequencies 
$\rho_{lat}(\omega)$ instead of the continuum expression (\ref{rho_tree_level}). 
Application of the function $\rho_{lat}(\omega)$ takes discretization effects in temporal direction into account  
and, as the result, considerably improves the quality of the fit. 

Evidently the inclusion of interactions modifies the tree level formula. 
However, due to the asymptotic freedom at large frequencies one can expect that this
modification is not dramatic. In particular, it is seen from (\ref{rho_NLO})
that one-loop radiative corrections modify 
the spectral function by a factor close to unity.  Inspired by this 
observation, in expression (\ref{rho1}) we multiplied the $\rho_{lat}(\omega)$
by some factor $A$ which effectively accounts radiative corrections 
at large frequencies. Our results show that the introduction of the factor $A$
is crucial for successfull description of the lattice data.

\begin{table}[]
\centering
\begin{tabular}{|c|c|c|c|c|c|c|c|}
\hline
                $T / T_c$ & fit type & $A$ & $\omega_0 / T$ & $\eta / s$ &  $\gamma$ & C & $\chi^2 / \mbox{dof}$ \\ \hline
\multirow{4}{*}{$0.90$}  & $\rho_1$ & 1.108(5) & 8.7(4) & 0.64(16) & -- & -- & 0.9 \\ \cline{2-7} 
                         & $\rho_2$ & 1.108(3) & 8.5(4) & 0.59(13) & 3.9(1.5) & -- & 1.2 \\ \cline{2-7}
                         & $\rho_3$ & 1.109(3) & 8.5(6) & 0.57(25) & -- & 0.002(53) & 1.2 \\ \hline
\multirow{4}{*}{$0.925$} & $\rho_1$ & 0.921(3) & 8.7(3) & 0.49(6) & -- & -- & 1.9 \\ \cline{2-7} 
                         & $\rho_2$ & 0.921(2) & 8.7(3) & 0.49(6) & 4.7(1.1) & -- & 2.1 \\ \cline{2-7}
                         & $\rho_3$ & 0.921(2) & 8.7(7) & 0.55(17) & -- & -0.003(19) & 2.1 \\ \hline
\multirow{4}{*}{$0.95$}  & $\rho_1$ & 0.942(5) & 7.7(5) & 0.22(9) & -- & -- & 1.0 \\ \cline{2-7} 
                         & $\rho_2$ & 0.940(4) & 7.6(5) & 0.20(8) & 2.5(1.3) & -- & 1.5 \\ \cline{2-7}
                         & $\rho_3$ & 0.940(8) & 7.6(7) & 0.24(15) & --& 0.002(37) & 1.5 \\ \hline
\multirow{4}{*}{$1.0$}   & $\rho_1$ & 0.998(13) & 7.3(5) & 0.23(3) & -- & -- & 1.8 \\ \cline{2-7} 
                         & $\rho_2$ & 0.998(13) & 7.3(5) & 0.21(5) & 3.7(8) & -- & 2.1 \\ \cline{2-7}
                         & $\rho_3$ & 0.998(13) & 7.3(1.1) & 0.24(8) & --& -0.007(65) & 2.1 \\ \hline
\multirow{4}{*}{$1.1$}   & $\rho_1$ & 0.927(8) & 7.0(7) & 0.18(5) & -- & -- & 1.3 \\ \cline{2-7} 
                         & $\rho_2$ & 0.927(8) & 6.9(6) & 0.17(4) & 4.1(1.4) & -- & 1.4 \\ \cline{2-7}
                         & $\rho_3$ & 0.927(8) & 7.2(1.0) & 0.15(5) & --& 0.02(3) & 1.4 \\ \hline
\multirow{4}{*}{$1.2$}   & $\rho_1$ & 0.819(8) & 7.6(4) & 0.21(3) & -- & -- & 1.6 \\ \cline{2-7} 
                         & $\rho_2$ & 0.818(7) & 7.6(5) & 0.21(3) & 5.4(8) & -- & 1.8 \\ \cline{2-7}
                         & $\rho_3$ & 0.818(9) & 7.6(5) & 0.22(6) & --& -0.004(28) & 1.8 \\ \hline
\multirow{4}{*}{$1.35$}  & $\rho_1$ & 0.932(8) & 7.7(5) & 0.22(3) & -- & -- & 0.9 \\ \cline{2-7} 
                         & $\rho_2$ & 0.932(8) & 7.7(5) & 0.22(3) & 2.3(1.0) & -- & 1.0 \\ \cline{2-7}
                         & $\rho_3$ & 0.932(8) & 7.9(1.0) & 0.20(7) & --& 0.01(5) & 1.0 \\ \hline
\multirow{4}{*}{$1.5$}   & $\rho_1$ & 0.932(9) & 9.0(4) & 0.28(2) & -- & -- & 1.0 \\ \cline{2-7} 
                         & $\rho_2$ & 0.932(9) & 9.0(4) & 0.27(2) & 2.6(7) & -- & 1.1 \\ \cline{2-7}
                         & $\rho_3$ & 0.932(9) & 9.1(4) & 0.27(7) & --& 0.002(31)  & 1.1 \\ \hline
\end{tabular}
\caption{The parameters of the functions $\rho_1(\omega)$, $\rho_2(\omega)$, $\rho_3(\omega)$ 
obtained from the fit of the lattice data. Instead of the parameter $B$
in the second column we show the ratio $\eta/s=\pi B/s$. }
\label{tab1}
\end{table}

A drawback of the ansatz (\ref{rho1}) is that this function has a discontinuity at the point 
$\omega=\omega_0$. It causes no difficulties to build the spectral function
with the properties similar to the ansatz (\ref{rho1}) but free from this drawback
\begin{equation}
\rho_2(\omega) = \frac 1 2 B T^3 \omega \left(1 + \tanh \left[ {\gamma (w_0 - w) } \right] \right) + 
\frac 1 2 A  \rho_{lat} ( \omega ) \left(1 + \tanh \left[ {\gamma (w - w_0) } \right] \right).
\label{rho2}
\end{equation}
Notice that in this ansatz we have introduced the parameter $\gamma$ which controls 
the width  of the transition from the infrared hydrodynamic regime  
to the ultraviolet regime of the asymptotic freedom in the spectral function. Evidently 
the width of the transition is $\sim 1/\gamma$. 

In the ansatzes (\ref{rho1}), (\ref{rho2}) the first-order hydrodynamic regime directly continues to the regime of the asymptotic freedom.
However, it is reasonable to assume that there is a region of frequencies where the spectral function 
deviates from the first-order hydrodynamics. In order to study this deviation 
 in addition to spectral functions (\ref{rho1}), (\ref{rho2}) we use the following ansatz
\begin{equation}
\rho_3(\omega) = B T^3 \omega \left(1 + C \omega^2 \right) \theta(\omega_0 - \omega) + 
 A  \rho_{lat} ( \omega ) \theta(\omega - \omega_0).
\label{rho3}
\end{equation}
In last formula we introduced a correction to the first-order hydrodynamic approximation which is controlled 
by the parameter $C$. We did not introduce next-to-leading order correction to the first-order hydrodynamics $\sim \omega^2$ since 
 the spectral function is an odd function of frequency.  
Notice that in the hard-thermal-loop framework the hydrodynamic behavior at small 
frequencies is replaced by the transport peak of a final width $\omega \to \omega/(1+\Gamma \omega^2)$ \cite{Aarts:2002cc}.
So, the ansatz (\ref{rho3}) can be considered as a first term of the expansion of the 
transport peak with $\Gamma=-C$.

We fit lattice data ($14 \geqslant x_0/a \geqslant 2$) for each temperature by the formula (\ref{spectr_corr}) 
with the spectral functions (\ref{rho1}), (\ref{rho2}), (\ref{rho3}). 
In the Table \ref{tab1} we show the parameters of functions (\ref{rho1}), (\ref{rho2}), (\ref{rho3}) obtained through this fit.
Instead of the parameter $B$ in the second column of Table \ref{tab1} we show the ratio $\eta/s$ which is related to the $B$
as $\eta/s=\pi B/s$. 

Now few comments are in order

\begin{itemize}
\item It is seen from Table \ref{tab1} that functions (\ref{rho1}), (\ref{rho2}), (\ref{rho3}) fit 
lattice data for various temperatures quite well.
It is also seen that for all ansatzes the ratio $\eta/s$ quickly drops 
when temperature approaches the critical point $T_c$ and then either slowly rises after $T_c$ or stays
constant. This behavior 
was seen in various models aimed at the calculation of shear viscosity in QCD. 

\item The values of $\eta/s$, $A$ and $\omega_0$ obtained through the 
fitting of the data by various ansatzes at the same temperature are in agreement with each other
within the uncertainty of the calculation. However, the 
uncertainties for these parameters are different for various ansatzes.  

\item The values of the threshold parameter $\omega_0$ for all ansatzes are physically well motivated. The value of the strong 
coupling constant at the threshold parameter $\omega_0$ ($\omega_0 \sim 2-3$ GeV in physical units) 
is $\alpha_s(\omega_0) \sim 0.2-0.3$. This allows us to expect that perturbative expression for the 
spectral function is applicable for $\omega > \omega_0$. The values of the factor $A$ for all temperatures, 
which takes into account radiative corrections, are close to unity what confirms applicability of 
the asymptotic freedom at high frequencies. 

\item Notice also that contrary to the infrared part of the spectral function the parameters of the 
ultraviolet part for all ansatzes are determined from the fit with 
a very good accuracy. This feature results from the fact that the dominant contribution 
to the correlation function is due to high frequencies.  

\item The ansatz $\rho_3(\omega)$ allows one to study the deviation from the first-order hydrodynamics. 
This deviation is controlled by the parameter $C$. From Table \ref{tab1} one sees 
that within the uncertainty of the calculation the values of the $C$ are zero for all temperatures. 
This fact implies that our data do not allow to observe the deviation from the 
first-order hydrodynamics.

\end{itemize}

It is worth to note that we tried to fit our data by the spectral
function similar to the $\rho_1(\omega)$ but with the substitution 
$\omega \to \omega/(1+\Gamma \omega^2)$. This substitution accounts 
for the transport peak \cite{Aarts:2002cc}. The result of this fit is 
very similar to that for ansatz (\ref{rho3}). The parameters $\Gamma$
equal zero within the uncertainty of the calculation.

Low frequency parts of spectral functions (\ref{rho1}) and (\ref{rho2}) 
are given by the first-order hydrodynamic expression $\sim \omega$. 
One can expect that the first-order hydrodynamic approximation works well up to $\omega \leqslant \pi T \simeq 1$ GeV \cite{Meyer:2008sn}. 
From the other side high frequency perturbative expression for the spectral function is fixed very 
accurately and it works well for $\omega \geqslant \omega_0 \sim 3$ GeV. The form of the spectral function 
in the region $1~\mbox{GeV} \leqslant \omega \leqslant 3~\mbox{GeV}$ is not clear. We believe that
poor knowledge of the spectral function in the region $1~\mbox{GeV} \leqslant \omega \leqslant 3~\mbox{GeV}$ 
is the main source of the uncertainty of the calculation based on the fitting of lattice data 
by the functions $\rho_1(\omega)$ and $\rho_2(\omega)$. Notice that this source of uncertainty in the values of shear viscosity is not accounted in 
Table \ref{tab1}.

The function $\rho_3(\omega)$ modifies the first-order hydrodynamic 
expression in the intermediate 
region due to the term $\sim C \omega^2$. Thus the function $\rho_3(\omega)$ at least partly 
takes into account uncertainty in shear viscosity due to our poor knowledge of the spectral function 
in the intermediate region. For this reason we take the results for the 
ratios $\eta/s$ obtained through the fitting by the function $\rho_3(\omega)$ as the 
results of this section. In addition to the statistical uncertainties in the ratio $\eta/s$ shown in Table \ref{tab1}, 
there are uncertainties in the entropy density $s$ and the renormalization coefficient
of the clover discretized energy-momentum tensor (\ref{clover}). The former uncertainties are $4-6$ \% for all 
temperatures under consideration. The latter uncertainties are $\sim 3$ \% for the temperatures $T/T_c \geqslant 1.0$,
$\sim 6$ \% for temperatures $T/T_c=0.925, 0.95$ and $\sim 12$ \% for the temperature $T/T_c=0.9$. 
The results for the ratios $\eta/s$ obtained within the fitting procedure including all uncertainties are shown 
in the second column of Table \ref{tab2} and in Fig.~\ref{fig:ratio_results}.

\section*{The Backus-Gilbert method for the spectral function}
In this section we are going to determine the ratio $\eta/s$ using the 
Backus-Gilbert(BG) method \cite{Backus:1, Backus:2}
\footnote{In QCD this approach was recently applied in papers \cite{Brandt:2015sxa, Brandt:2015aqk}. 
Backus-Gilbert method was also recently applied in the lattice calculation of graphene conductivity \cite{Boyda:2016emg}.}. 
This approach has considerable advantage over the 
method based on the fitting procedure: one does not need to know the parametrical form of the spectral function to carry out the calculation.

The method can be formulated as follows\footnote{Here we follow the designations of \cite{Brandt:2015sxa, Brandt:2015aqk}}. 
One needs to solve equation (\ref{spectr_corr}). To do this we rewrite it in the following form
\beq
C(x_i)=\frac 1 {T^5} \int_0^{\infty} d \omega \frac {\rho(\omega)} {f(\omega)} K(x_i,\omega), 
\eeq
where $f(\omega)$ is some function with the property $f(t)|_{t \to 0} \sim t$,
$x_i$ are lattice points where the calculation of the $C(x_i)$ are carried out and $K(x, \omega)$ is a rescaled kernel of the integral equation
\beq
K(x, \omega) = f(\omega) \frac {\cosh \omega \biggl(\frac 1 {2T} - x \biggr )} {\sinh \frac {\omega} {2 T}}, 
\eeq 
Our aim is to determine the $\rho(\omega)$. In the BG method instead of the function $\rho(\omega)$ we consider 
the estimator of this function $\bar \rho(\bar \omega)$ which can be written as 
\beq
\bar \rho(\bar \omega) = f(\bar \omega) \int_0^{\infty} d \omega \delta (\bar \omega, \omega) \frac {\rho(\omega)} {f(\omega)}, 
\label{barf}
\eeq
where the function $\delta(\bar \omega, \omega)$ is called the resolution function. This 
function has a peak around the point $\bar \omega$ and it is normalized as 
$\int_0^{\infty} d \omega \delta(\bar \omega, \omega)=1$. The function is expanded over the $K(x_i, \omega)$ as
\beq
\delta(\bar \omega, \omega) = \sum_i q_i(\bar \omega) K(x_i, \omega).
\eeq
For this resolution function the estimator is a linear conbination of the values of the correlation function
\beq
\bar \rho(\bar \omega) = f(\bar \omega) \sum_i q_i(\bar \omega) C(x_i)
\label{barrho} 
\eeq
Evidently to get a better approximation for the $\rho(\omega)$ by the estimator $\bar \rho(\bar \omega)$ one 
should minimize the width of the $\delta(\bar \omega, \omega)$. However, a very narrow peak might build an estimator fitting the points themselves, but not the physics (generality) they present. This means that any method of this kind should be regularized.

The Backus-Gilbert is aimed at minimization of the Backus-Gilbert functional $\displaystyle \mathcal{H}(\rho(\omega)) = \lambda \mathcal{A}(\rho(\omega)) + (1 - \lambda) \mathcal{B}(\rho(\omega))$. 
The component $\mathcal{A}$ represents the width of the resolution function (the second moment of distribution): $ \displaystyle \mathcal{A} = \int_0^{\infty} d \omega \delta(\bar \omega, \omega) (\omega - \bar \omega)^2$. In principle, it could be any other function with the same physical meaning. The advandate of the second moment is that it is quadratic in $\omega$ and $\bar \omega$, making analytical minimization possible.

The component $\mathcal{B}(\rho(\omega)) = \mbox{Var}[\rho(\omega)]$ punishes $\rho(\omega)$ for being too dependent on the data. In terms of the covariance matrix and $q$--functions, it reads $\displaystyle \mathcal{B}(\vec{q}) = \vec{q}^T \hat{S} \vec{q}$.

Putting everything together, the minimization of the $\mathcal{H}$ functional gives the following values of the coefficients
\beq
q_i(\omega) &=& \frac { \sum_j W^{-1}_{ij} (\bar \omega) R(x_j) } { \sum_{kj} R(x_k) W^{-1}_{kj} (\bar \omega) R(x_j) }, \\
W_{ij}(\bar \omega) &=& \lambda\int_{0}^{\infty} d \omega K(x_i,\omega) (\omega- \bar \omega)^2 K(x_j,\omega) + (1 - \lambda) S_{i j}, \\
R(x_i) &=& \int_0^{\infty} d \omega K(x_i,\omega). 
\eeq
If the $\lambda$ is close to $1$, the resolution function has the smallest width and 
the estimator gives the best approximation for the spectral function. 
However, the application of the Backus-Gilbert method with $\lambda \sim 1$
to the calculation of shear viscosity gives rise to large uncertainties. The result becomes very dependent on the data, the spectral function turns out to be noisy and unstable.
Statistical uncertainties can be improved at the expense of increasing 
the width of the resolution function through decreasing the value of 
the $\lambda$--parameter. 

\begin{figure}[t]
\begin{center}
\includegraphics[scale=0.5,angle=0]{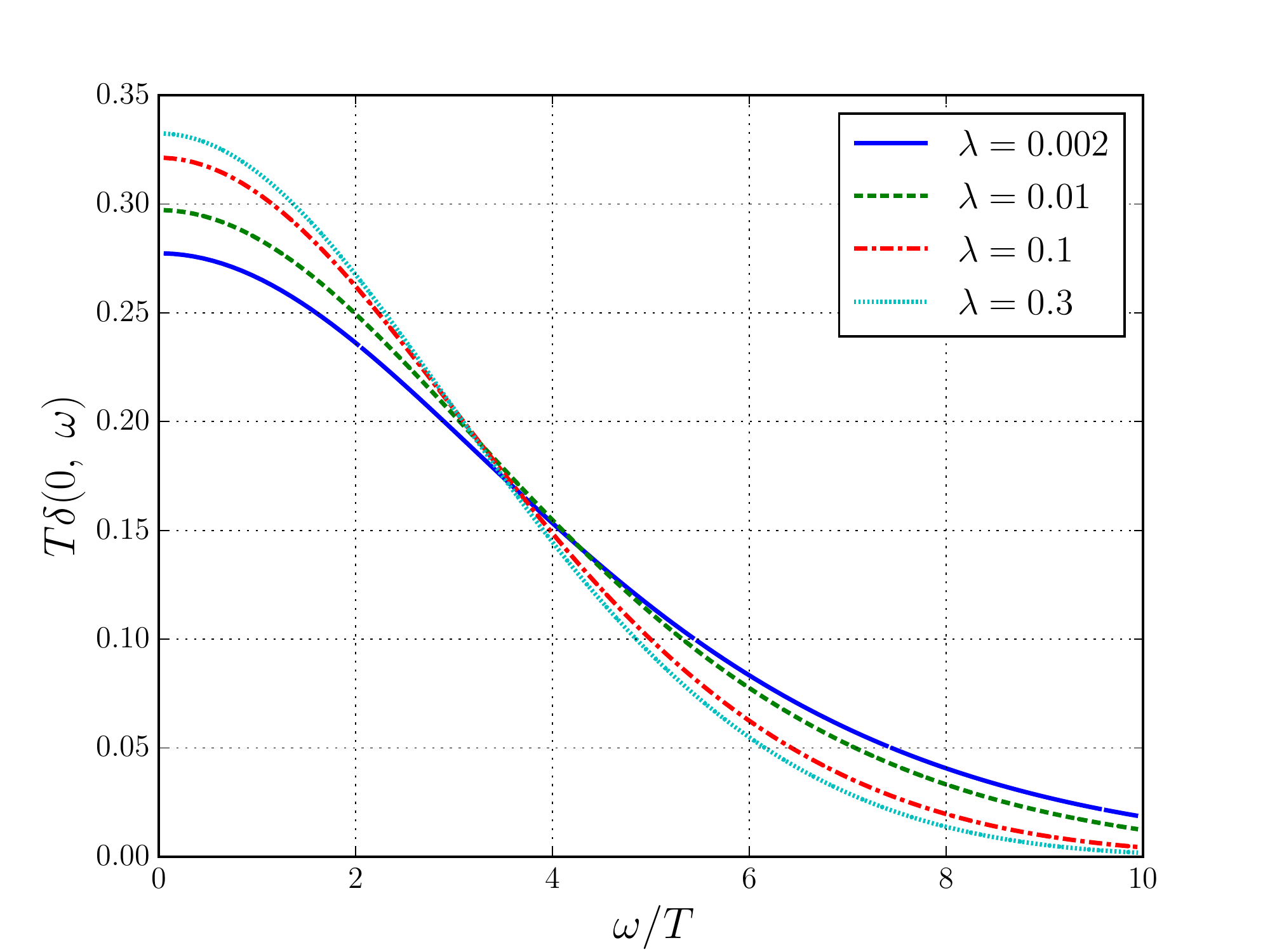}
\caption{The resolution function at temperature $T/T_c=1.35$ and $\bar \omega=0$ for various values of the $\lambda$.}
\label{fig:resolution}
\end{center}
\end{figure}

Now let us discuss the choice of the function $f(x)$. 
To determine the spectral density at small frequencies it is
reasonable to choose the following function
\beq
f_1(x)=x
\eeq
In this case the ratio $\displaystyle\rho(\omega)/f_1(\omega)|_{\omega \to 0} = \eta / \pi$. 
The main motivation for choosing this function is that it 
gives rather small width of the resolution function at small frequencies. 
   
To study the spectral function at large frequencies we choose the following function 
\beq
f_2(x)=\frac {\rho_{lat}(x)} {(\tanh(x/4T))^2} 
\eeq
One can expect that due to asymptotic freedom 
at large frequencies $\omega \gg \Lambda_{QCD}$ the ratio $\rho(\omega)/f_2(\omega)$ 
behaves like constant. The width of the resolution function with the $f_2(x)$ is 
 larger than that with the $f_1(x)$.       
\begin{figure}[t]
\includegraphics[scale=0.4,angle=0]{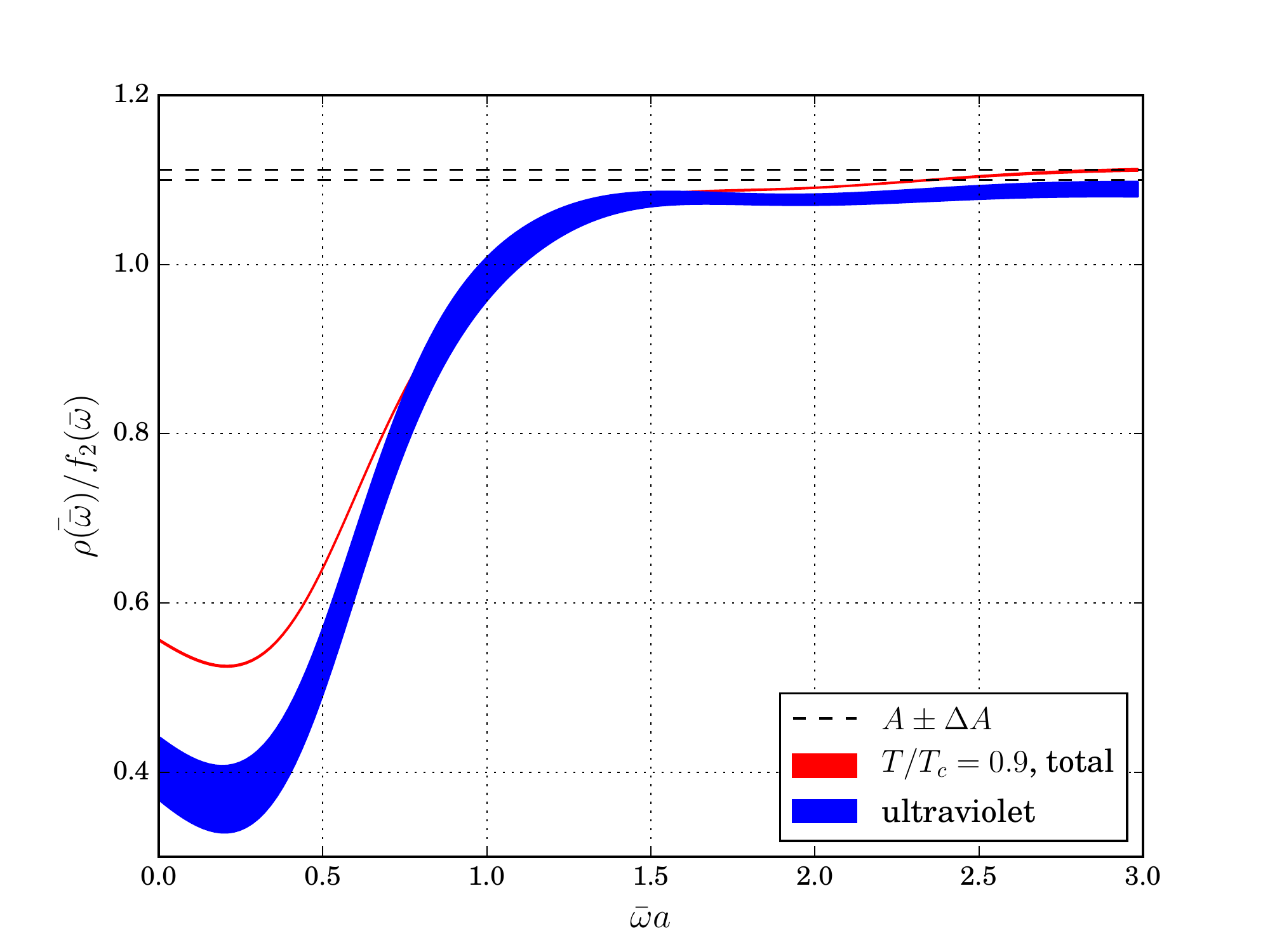}
\includegraphics[scale=0.4,angle=0]{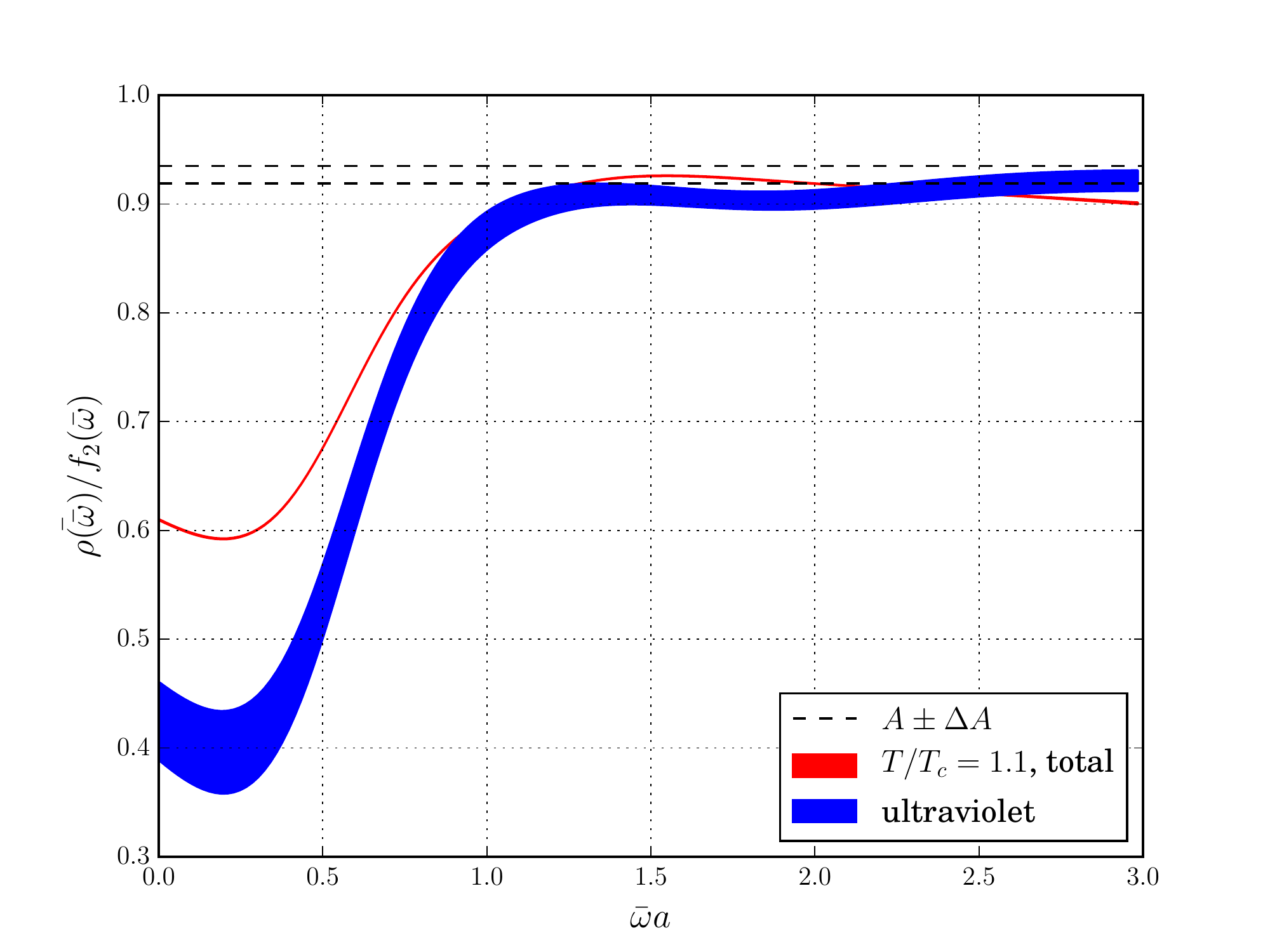} \\
\includegraphics[scale=0.4,angle=0]{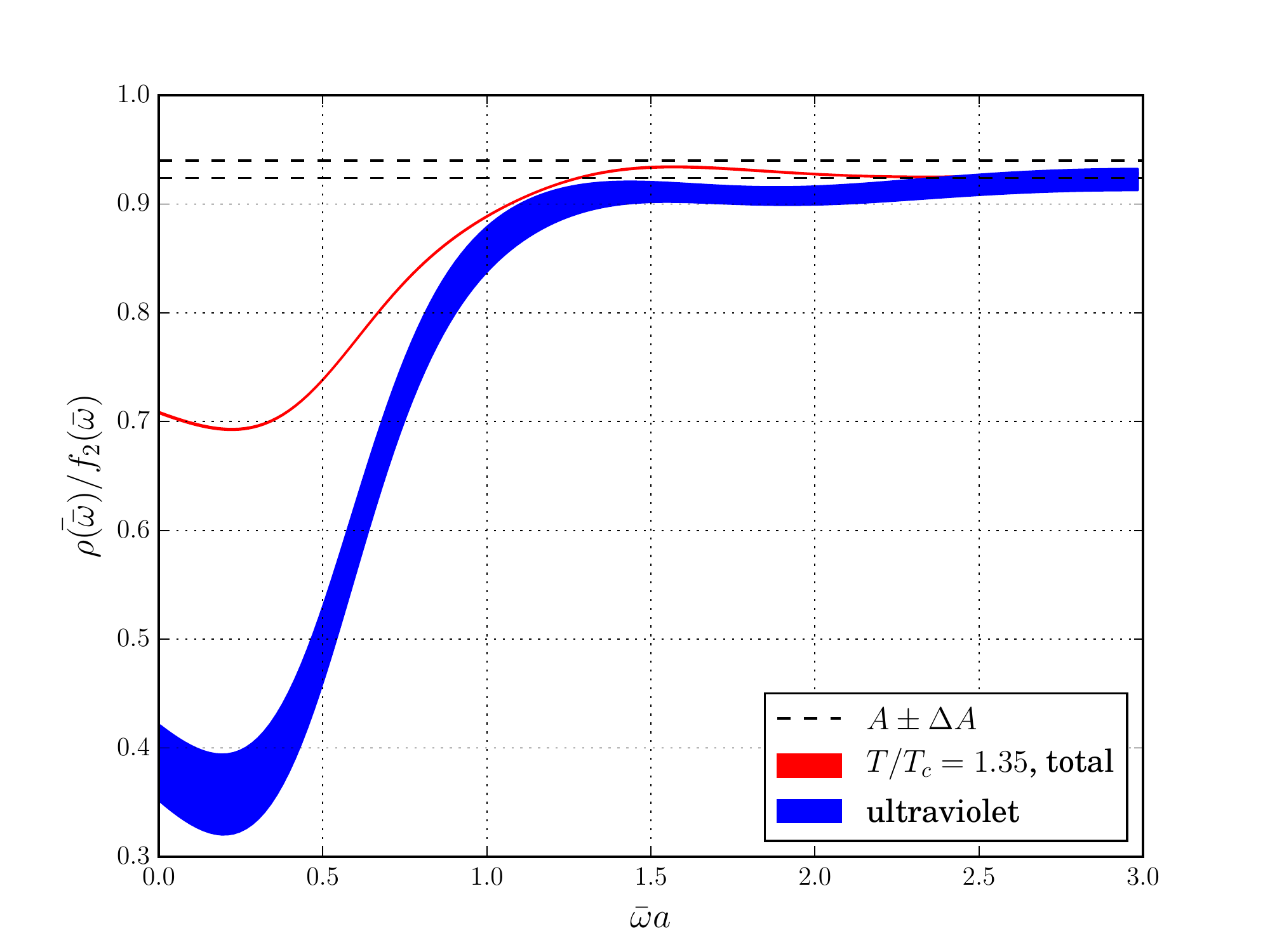}
\includegraphics[scale=0.4,angle=0]{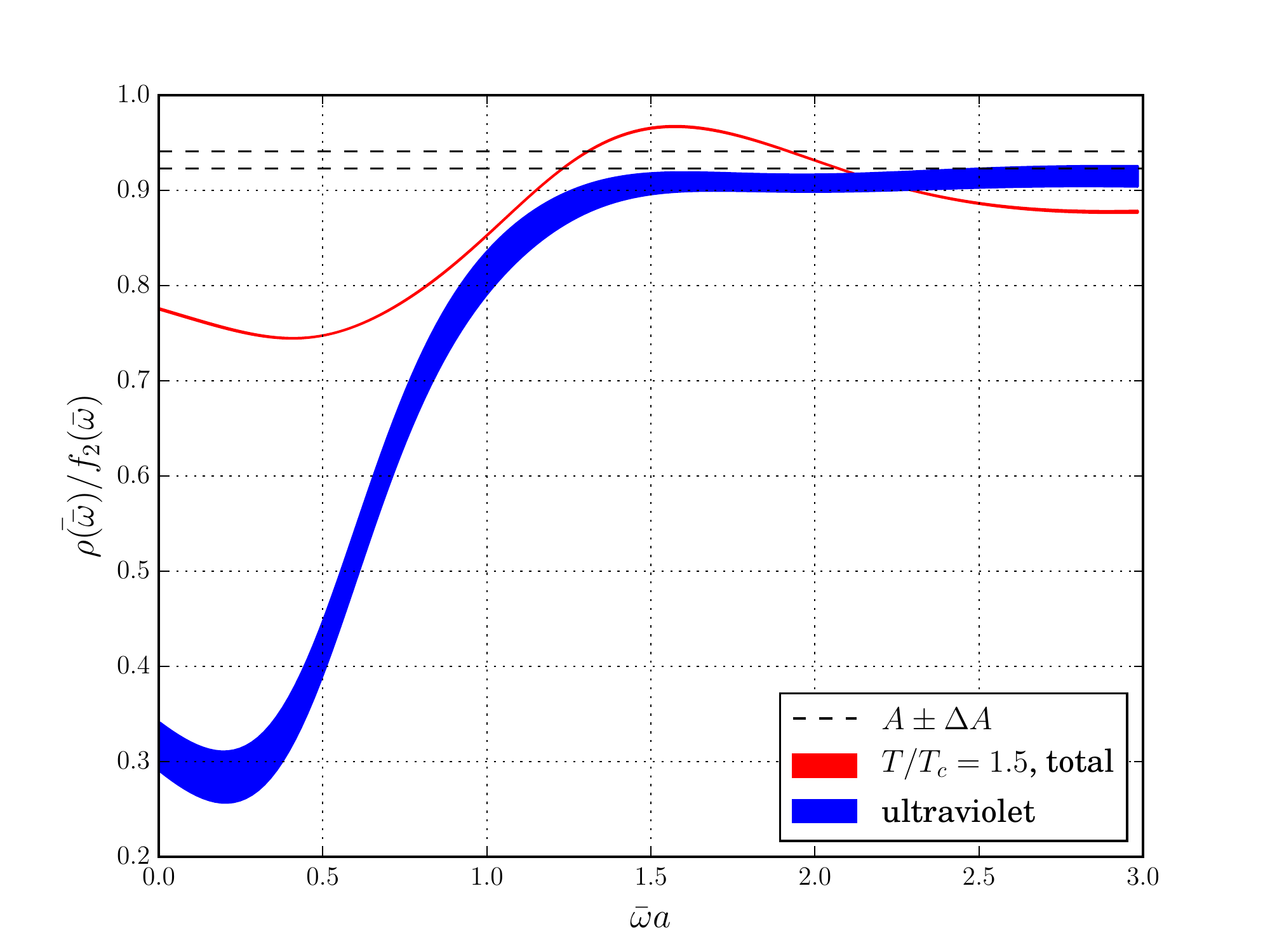}
\caption{The ratios $\bar \rho(\bar \omega)/f_2(\bar \omega)$ as a function of $\bar \omega a$ 
for the temperatures $T/T_c=0.9, 1.1, 1.35, 1.5$. 
Red curves correspond to spectral functions restored by the BG method from the data. 
Blue curves correspond to the ultraviolet contribution in the form (\ref{ultr}) convoluted with the resolution 
function. 
Dashed lines are values of the constants $A$ with uncertainties obtained within fitting procedure. }
\label{fig:bar_rho}
\end{figure}

Now let us proceed to the calculation of shear viscosity. To do this 
we use the function $f_1(x)$. As to the $\lambda$-parameter 
we choose $\lambda=0.002$. For this value of the $\lambda$ the uncertainty in the restored 
spectral density at zero frequency is smaller than 1\% for all temperatures under consideration. 
In Fig.~\ref{fig:resolution} we plot the resolution function at temperature $T/T_c=1.35$ and $\bar \omega=0$ for 
various values of the $\lambda$. The resolution functions for the other temperatures 
are very similar to those in Fig.~\ref{fig:resolution} and we do not show them here.  

From Fig.~\ref{fig:resolution} one sees that the width of the resolution function 
at $\lambda=0.002$ is $\Delta \omega \sim 4T$. If for a while we forget about 
the ultraviolet contribution, the convolution of the spectral function $\rho(\omega)$ with 
the resolution function (\ref{barf}) gives some average of the spectral function 
over the interval with the width $\sim 4T$. One can expect that the first-order hydrodynamic 
approximation works well up to $\omega \leqslant \pi T \simeq 1$ GeV \cite{Meyer:2008sn},
what covers most of the interval $(0,4T)$

Now let us discuss the ultraviolet contribution to convolution (\ref{barf}). 
According to the results of previous section the ultraviolet part of the spectral 
function starts to work 
for frequencies $\omega/T \sim 7-8$. From Fig.~\ref{fig:resolution} one sees that 
the resolution function is considerably suppressed in this region. However, it 
is not possible to disregard ultraviolet contribution since the spectral function at large 
frequencies rises very quickly $\rho(\omega) \sim \omega^4$. 
Calculation shows that for the most temperatures under consideration 
the contribution which results from the ultraviolet part of the spectral function 
is larger than the contribution of hydrodynamic part of the spectral function. 
So, to get reliable result for shear viscosity one should subtract the ultraviolet contribution. 
\begin{figure}[t]
\includegraphics[scale=0.6,angle=0]{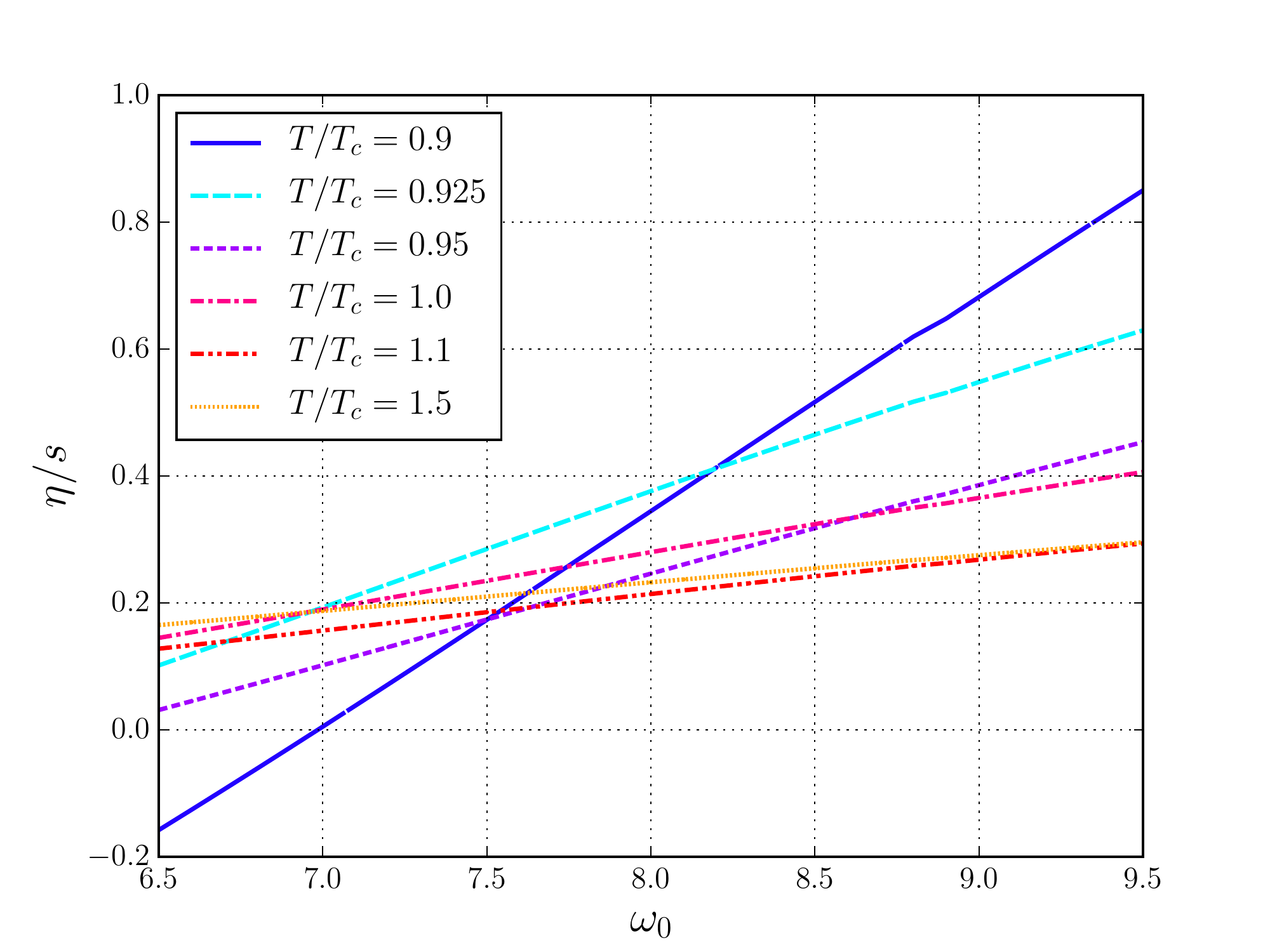}
\caption{The ratio $\eta/s$ with the ultraviolet contribution subtracted as a function of $\omega_0$
for the temperatures $T/T_c=0.9, 0.925, 0.95, 1.0, 1.1, 1.5$. 
The parameter $A$ is taken as the central value of the fit of lattice data 
by the function $\rho_3(\omega)$.
 }
\label{fig:ratio_w0}
\end{figure}

To study the spectral function at large frequencies we are going to use the BG method 
with the function $f_2(\omega)$ and $\lambda=0.002$. In Fig.~\ref{fig:bar_rho} 
we plot the ratios $\bar \rho(\bar \omega)/f_2(\bar \omega)$ as a function 
of $\bar \omega a$ for the temperatures $T/T_c=0.9, 1.1, 1.35, 1.5$. For the 
other temperatures under consideration the figures are similar and we do not show them here. 
Red curves correspond to the spectral functions restored by the BG method from our data.
In order to compare the results of the BG method with the results obtained in the previous section
we took the spectral function at large frequencies 
\beq
\rho_{ult}(\omega)=A \rho_{lat}(\omega) \theta(\omega-\omega_0),
\label{ultr}
\eeq
and convoluted it with the resolution function. The values of the $A$ and $\omega$ with uncertainties
were taken from the fitting procedure (see Table \ref{tab1}). The results 
are presented as blue curves. Finally we plotted 
dashed lines which correspond to the values of the constants $A$ with uncertainties obtained 
within fitting procedure.

Now few comments are in order
\begin{itemize}

\item It is seen from Fig.~\ref{fig:bar_rho} that the red curves can be separated into 
two parts. The first part is the spectral function for small frequencies  $\bar \omega a \leqslant 0.5$ ($\bar \omega T \leqslant  8$). 
One can say that the spectral function in this region is in the infrared regime. 
After $\bar \omega a \sim  0.5$ ($\bar \omega T \sim  8$) there is a transition to the second regime where the spectral 
function is close to the ultraviolet asymptotic which is given by the constant $A$.   

\item It is also clear from Fig.~\ref{fig:bar_rho} that the behavior of the blue curves which represents 
the ultraviolet contribution to the ratio $\bar \rho(\bar \omega)/f_2(\bar \omega)$ is 
similar to the red ones. In the ultraviolet regime the blue and red curves are close to 
each other. Transition from the ultraviolet to the infrared regime takes place within the same 
region in $\bar \omega a$. 

\item In the infrared region the red curves are higher than the blue ones. The difference 
between them can be attributed to contribution of the spectral functions at 
small frequencies. One sees that the smaller the temperature the smaller the difference. 
If we recall that shear viscosity is related to the spectral function at small 
frequencies one can state that shear viscosity drops with temperature. Our results 
assume that the entropy density $s$ drops with temperature more quickly than shear viscosity. 
As a result the ratio $\eta/s$ rises below $T_c$. 

\item From Fig.~\ref{fig:bar_rho} one can see that in the ultraviolet region the restored spectral 
function is not a constant but some slowly varying function of the $\bar \omega a$.
The deviation of this function from the asymptotic value $A$ obtained within fitting procedure 
is very small for all temperatures. For the most temperatures the deviation is few percent.  
The deviation of the restored spectral function from the asymptotic value $A$ can be attributed 
to radiative corrections to the tree level spectral function which, evidently, more complicated 
than constant.   

\end{itemize}

The study carried out in this section  allows us to state that formula (\ref{ultr}) describes ultraviolet part of 
the spectral function quite well. For this reason below we are going to use (\ref{ultr}) as a 
model for the ultraviolet part of the spectral function. The value of the constant $A$
will be determined from the variation of the restored ratio $\bar \rho(\bar \omega)/f_2(\bar \omega)$ 
in the region $\bar \omega a \in (1.5, 3)$. This interval is chosen since  
the contribution of the infrared part of the spectral function is this region is small 
and the ratios $\bar \rho(\bar \omega)/f_2(\bar \omega)$ for all temperatures are in the 
ultraviolet regime. The values of the constants $A$ determined in this way are in agreement 
with that obtained within fitting procedure.

In addition to the constant $A$ the ultraviolet part (\ref{ultr}) depends on the 
threshold parameters $\omega_0$. Thus, if one subtracts the ultraviolet contribution 
in the form (\ref{ultr}), the ratio $\eta/s$ obtained within the BG method will depend on the value of $\omega_0$. 
To study this dependence in Fig.~\ref{fig:ratio_w0} we plot the ratio $\eta/s$ as a function of $\omega_0$
 for the temperatures $T/T_c=0.9, 0.925, 0.95, 1.0, 1.1, 1.5$.
The curves for the temperatures $T/T_c=1.2, 1.35$ are very close to the curve at $T/T_c=1.5$.
For this reason we do not show these temperatures on the figure. 
The parameter $A$ is taken at the central value of the fit of lattice data 
by the function $\rho_3(\omega)$ (see Table \ref{tab1}). From Fig.~\ref{fig:ratio_w0} one sees that 
the larger the temperature the larger the slope of the curves and the weaker the dependence of the $\eta/s$ 
on $\omega_0$. The dependence of the $\eta/s$ on $\omega_0$ is weak for the temperatures $T/T_c \geqslant 1.0$
and it is stronger for the $T/T_c<1.0$. The strongest dependence of the $\eta/s$ on $\omega_0$ is for the 
temperature $T/T_c=0.9$. 
We believe that this property stems from the already mentioned fact: shear viscosity of gluodynamics drops  
with temperature and the extraction of viscosity from the observable which contains 
large ultraviolet contribution becomes more and more complicated for lower temperatures.   
 
Unfortunately it is not quite clear how the value of the threshold parameter $\omega_0$
can be determined within the BG method. Note, however, that the position of the 
transition from the infrared to ultraviolet regime (see Fig.~\ref{fig:bar_rho}) coincide for both restored spectral function
and for the function (\ref{ultr}) with $\omega_0$ obtained within fitting procedure.
Note also that the values of the parameter $A$ obtained within the BG method and fitting predure
agrees quite well. For this reason one can expect that fitting procedure gives a good approximation 
for the value of $\omega_0$ and we will take it for the model of the ultraviolet 
contribution. 

\begin{figure}[t]
\includegraphics[scale=0.6,angle=0]{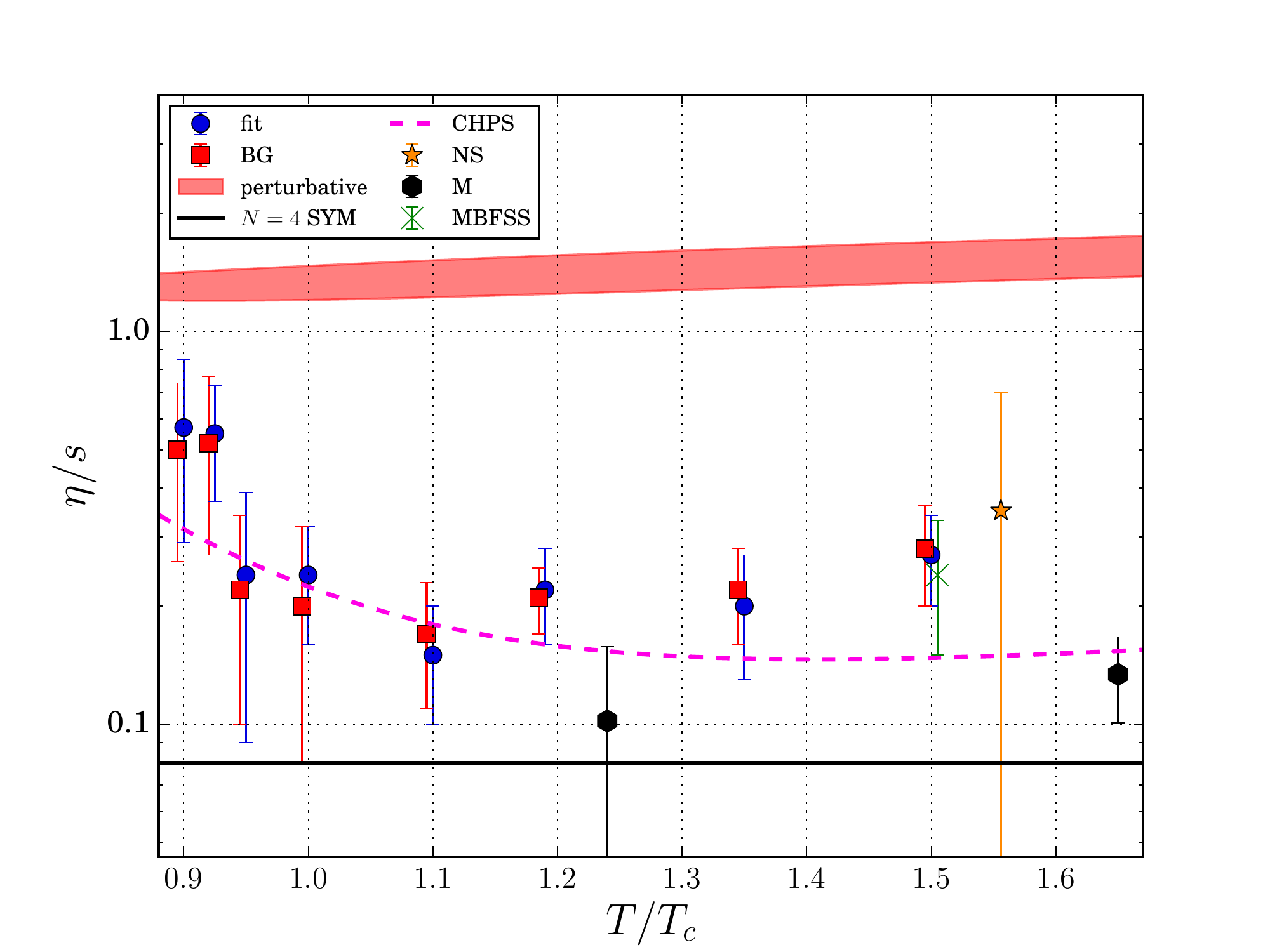}
\caption{The ratio $\eta/s$ in gluodynamics for various temperatures. 
The circle blue points correspond to the results obtained within fitting procedure and 
the square red points correspond to the BG method. We also plot the previous lattice 
results obtained by A.~Nakamura, S.~Sakai in \cite{Nakamura:2004sy} (the yellow star point), 
by H.Meyer in \cite{Meyer:2007ic} (the black hexagonal points) and by S.~Mages, S.~Borsnyi, Z.~Fodor, A.~Schfer, K.~Szab
in \cite{Mages:2015rea} (the green cross point). 
In addition we plot the result of $N = 4$ SYM theory at strong coupling $\eta/s = 1/4\pi$ \cite{Policastro:2001yc} (black solid line),
the result of perturbative calculation of the $\eta$ \cite{Arnold:2003zc} (red region) and
the result obtained by N.~Christiansen, M.~Haas, J.M.~Pawlowski, N.~Strodthoff in paper \cite{Christiansen:2014ypa} 
(violet curve). 
}
\label{fig:ratio_results}
\end{figure}

Subtracting the ultraviolet contribution from the ratio $\eta/s$ in the form (\ref{ultr})
we get the results of this section. These results are shown 
in the third column of Table \ref{tab2} and in Fig.~\ref{fig:ratio_results}. The uncertainties in Table \ref{tab2} 
and in Fig.~\ref{fig:ratio_results} are due to the uncertainties in the $A$, $\omega_0$, the entropy density $s$ 
and the renormalization constant of the energy-momentum tensor (see previous section).
From Table \ref{tab2} and Fig.~\ref{fig:ratio_results} one sees that the results obtained within two approaches applied in this paper agree with each other.  

In addition to the results obtained in this paper in Fig.~\ref{fig:ratio_results} 
we plot lattice results obtained in papers \cite{Nakamura:2004sy, Meyer:2007ic, Mages:2015rea}.
It is seen that our results are in agreement with the previous lattice studies.  

It is also interesting to draw the results of paper \cite{Christiansen:2014ypa}. In this paper 
the authors calculated the shear viscosity in Yang-Mills theory 
using an exact diagrammatic representation in terms of full propagators and vertices using gluon
spectral functions as external input. Our results are in agreement with the results of this paper.

In Fig.~\ref{fig:ratio_results} we also plot the value of the ratio $\eta/s$ for the $N = 4$ 
SYM theory at strong coupling $\eta/s = 1/4\pi$ \cite{Policastro:2001yc} and 
the results of perturbative calculation of the $\eta/s$. Perturbative results were obtained as 
follows. The scale $\Lambda$ for the running coupling constant in 
gluodynamics was taken from \cite{Capitani:1998mq}. 
The entropy density $s$ was taken at one-loop accuracy \cite{Kapusta:2006}. 
We took perturbative results for shear viscosity at next-to-leading-log approximation from paper \cite{Arnold:2003zc}.
In order to estimate the uncertainty of the perturbative results we varied the scale 
in the region from the first to the second Matsubara frequency $\mu \in (2 \pi T, 4 \pi T)$.
Comparing our results with other approaches 
one can conclude that within the temperature
range $T/T_c \in [0.9, 1.5]$ SU(3)--gluodynamics reveals the properties of a strongly interacting system,
which cannot be described perturbatively, and has the ratio $\eta/s$ close to the value $1/4\pi$ in $N=4$
Supersymmetric Yang-Mills theory.

\section*{Discussion and conclusion}

This paper is aimed at studying the temperature dependence of shear viscosity of SU(3)--gluodynamics
within lattice simulation. In particular, we measured the correlation functions of 
energy-momentum tensor $\langle T_{12}(x_0) T_{12}(0) \rangle$ 
at lattice $16\times32^3$ for the temperatures in the range $T/T_c\in[0.9, 1.5]$. 
In order to get small uncertainties in our results we used two-level algorithm which allowed us to reach the accuracy
not larger than $\sim 2-3\%$ at the distance $T x_0 = 0.5$ for all temperatures under 
consideration. For the other points the accuracy is better.

Using lattice data for the correlation functions we calculated the ratios $\eta/s$ for the temperatures 
under consideration. To do this we used physically motivated ansatzes for the spectral 
function with unknown parameters. These papameters were determined through the fitting procedure. 
All ansatzes used in this paper are different ways of the interpolation between 
hydrodynamic behavior at small frequencies and asymptotic freedom at large frequencies. 
These ansatzes fit lattice data quite well for all temperatures.  
Another approach used to calculate the ratio  $\eta/s$ is the Backus-Gilbert 
method. 

\begin{table}[h!]
\centering
\label{my-label}
\begin{tabular}{|c|c|c|c|c|}
\hline
$T / T_c$ & $\eta / s$, Gluodynamics  & $\eta / s$, Gluodynamics& $\eta / s$, $QCD_{N_f = 3}$ \\ 
 &  fitting procedure   & BG method  &  \\ \hline
0.90 & 0.57(28) & 0.50(24) & 0.59(28) \\ \hline
0.925 & 0.55(18) & 0.52(25) & 0.61(30) \\ \hline
0.95 & 0.24(15) & 0.22(12) & 0.26(14) \\ \hline
1.0 & 0.24(8) & 0.20(12) & 0.24(14) \\ \hline
1.1 & 0.15(5) & 0.17(6) & 0.21(7) \\ \hline
1.2 & 0.22(6) & 0.21(4) & 0.26(5) \\ \hline
1.35 & 0.20(7) & 0.22(6) & 0.28(7) \\ \hline
1.5 & 0.27(7) & 0.28(8) & 0.36(10) \\ \hline
\end{tabular}
\caption{The ratio $\eta/s$ for various temperatures obtained by various methods. 
The results for gluodynamics 
obtained using the fitting procedure are presented in the second column.
The results for gluodynamics 
obtained within the BG method are presented in the third column.
Estimation of the ratio $\eta/s$ for QCD with dynamical $N_f=3$  quarks are 
presented in the forth column.}
\label{tab2}
\end{table}

In Table \ref{tab2} and Fig.~\ref{fig:ratio_results} we plot the results obtained in this paper. 
From Table \ref{tab2} and Fig.~\ref{fig:ratio_results} one sees that the results obtained within 
two approaches applied in this paper agree with each other.

In addition in Fig.~\ref{fig:ratio_results} 
we plot lattice results obtained in papers \cite{Nakamura:2004sy, Meyer:2007ic, Mages:2015rea}.
It is seen that our results are in agreement with the previous lattice studies.  
In Fig.~\ref{fig:ratio_results} we also plot the value of the ratio $\eta/s$ for the $N = 4$ SYM theory at strong coupling $\eta/s = 1/4\pi$ 
and the results of perturbative calculation of the $\eta/s$. 
Comparing our results with other approaches one can conclude that the ratio $\eta/s$ for the gluodynamics is very close to 
$N = 4$ SYM and cannot be described perturbatively. 

It is also interesting to mention the results of paper \cite{Christiansen:2014ypa}. In this paper 
the authors calculated the shear viscosity in Yang-Mills theory 
using an exact diagrammatic representation in terms of full propagators and vertices using gluon
spectral functions as external input. Our results are in agreement with the results of this paper.

\begin{figure}[t]
\includegraphics[scale=0.6,angle=0]{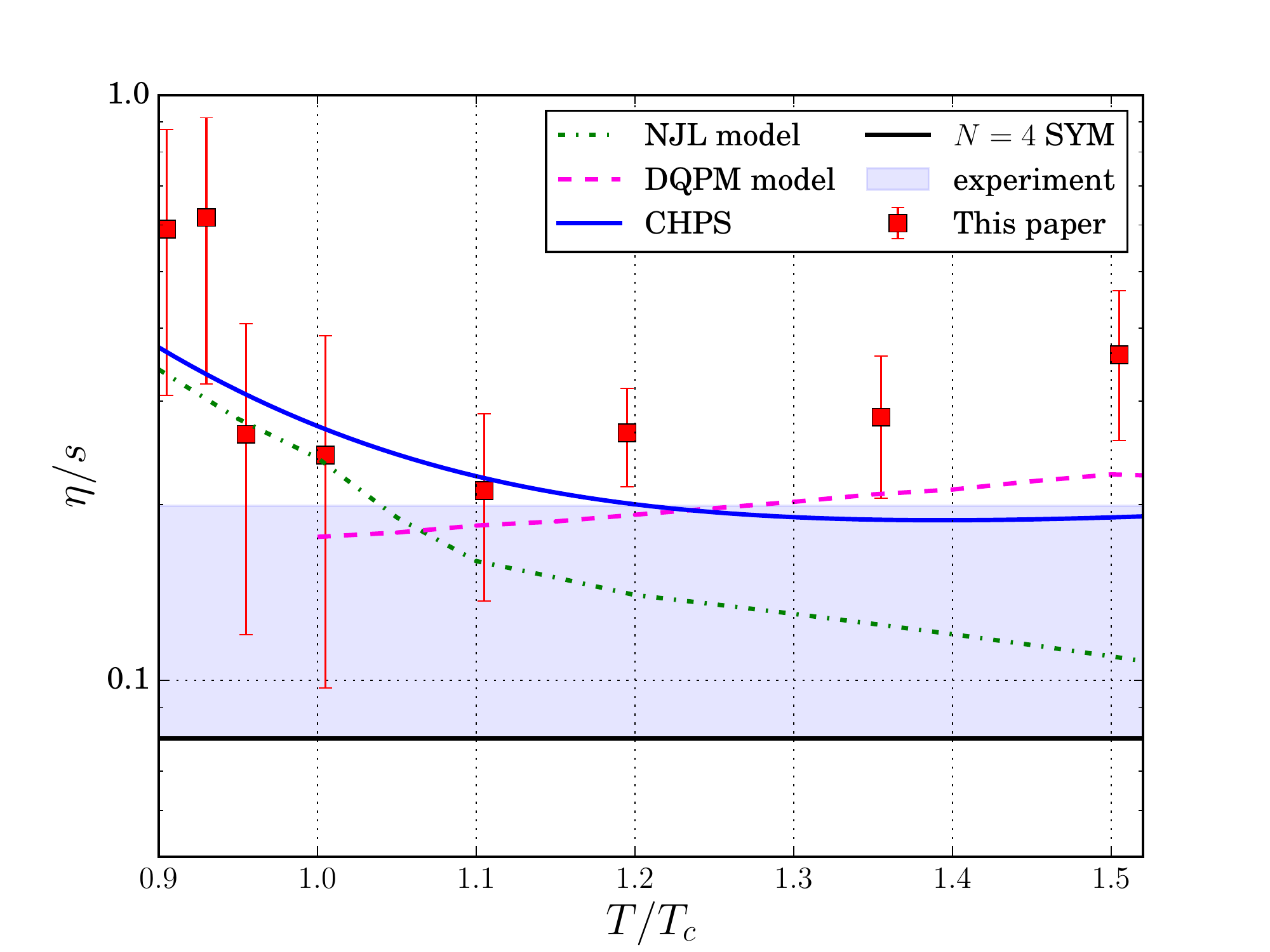}
\caption{The ratio $\eta/s$ in QCD with for various temperatures. 
The red square points represent the estimation of the $\eta/s$ in QCD
done in this paper. The green curve is the result of NJL model \cite{Marty:2013ita},
the violet curve is the result of dynamical quasiparticle approach \cite{Berrehrah:2016vzw}, 
blue curve is the result of paper \cite{Christiansen:2014ypa} and black line is the result of $N = 4$ SYM.
In addition we plot the grey region which shows the experimental bound on the ratio $\eta/s$ 
found from experiment data \cite{Song:2012ua}.
}
\label{fig:ratio_QCD}
\end{figure}

Today it is not possible to carry out lattice calculation of shear viscosity in QCD with dynamical fermions. 
However, one can estimate the ratio $\eta/s$ using the following formula
\beq
\displaystyle
\left( \eta / s \right)_{QCD} = \frac { \left(\eta/s \right)_{QCD} } { \left(\eta/s \right)_{YM} } 
\times \left( \eta/s \right)_{YM}
\eeq
The ratio $\bigl ( {\eta} / {s}  \bigr )_{YM}$ is calculated in this paper, while the  
ratio  $\displaystyle \left( \frac {\eta} {s} \right)_{\mbox{QCD}}  \bigg/  \left( \frac {\eta} {s} \right)_{\mbox{YM}}$ for $N_f=3$ quarks 
was estimated in paper \cite{Christiansen:2014ypa}. For the $\bigl ( {\eta} / {s}  \bigr )_{YM}$ 
we took the result obtained using the BG method. Our results for the ratio 
$\bigl ( {\eta} / {s}  \bigr )_{QCD}$  are shown in Table \ref{tab2} and in Fig.~\ref{fig:ratio_QCD}.
In addition in Fig.~\ref{fig:ratio_QCD} we plot the estimation of the 
$\eta/s$ obtained within various models: NJL \cite{Marty:2013ita},
dynamical quasiparticle approach \cite{Berrehrah:2016vzw}, the result of paper \cite{Christiansen:2014ypa}
and $N = 4$ SYM. Finally in Fig.~\ref{fig:ratio_QCD} we plot the gray region which shows the experimental bound on the ratio $\eta/s$ 
found from experiment data \cite{Song:2012ua}. It is seen that the results obtained in this paper agree with experiment.

\section*{Acknowledgements}
We would like to thank Anthony Francis who explained us how the BG method can be used to study of the utraviolet behavior of the spectral function. We also thank Alexander Nikolaev for valuable comments about the manuscript. Numerical  simulations  were  performed  at  the supercomputer  of  ITEP,  at  the  federal  center  for
collective  usage  at  NRC  ``Kurchatov  Institute'' (http://computing.kiae.ru/) and at MSU supercomputer
``Lomonosov''.  The work of AYK  was supported by RFBR grant 16-32-00048.  
The work of VVB, which consisted of calculation of shear viscosity within the BG approach, was supported by
the RSF grant under contract 15-12-20008.

\end{document}